\documentstyle[12pt]{article}
\newcommand{\resection}[1]{\setcounter{equation}{0}\section{#1}}

\begin{document}
\newcommand{\be}{\begin{equation}}
\newcommand{\ee}{\end{equation}}
\newcommand{\bea}{\begin{eqnarray}}
\newcommand{\eea}{\end{eqnarray}}
\newcommand{\nn}{\nonumber}
\newcommand{\dd}{\displaystyle}
\newcommand{\FF}{{\cal F}}
\newcommand{\AAA}{{\cal A}}
\newcommand{\CC}{{\cal C}}
\def\bra#1{\langle #1 |}
\def\ket#1{| #1\rangle }
\vspace*{4cm}
\begin{center}
  \begin{Large}
  \begin{bf}
INTEGRATION OVER A GENERIC\\ ALGEBRA\\
  \end{bf}
  \end{Large}
\end{center}
  \vspace{5mm}
\begin{center}
  \begin{large}
R. Casalbuoni\\
  \end{large}
Dipartimento di Fisica, Universita' di Firenze\\ I.N.F.N., Sezione di
Firenze\\
\end{center}
  \vspace{5mm}
  \vspace{2cm}
\vspace{6cm}
\begin{center}
Firenze Preprint - DFF - 270/02/1997
\end{center}
\newpage
\thispagestyle{empty}
\begin{quotation}
\vspace*{5cm}
\begin{center}
  \begin{bf}
  ABSTRACT
  \end{bf}
\end{center}
  \vspace{5mm}
\noindent
In this paper we consider the problem of quantizing theories
defined over configuration spaces described by non-commuting
parameters. If one tries to do that by generalizing the
path-integral formalism, the first problem one has to deal with is
the definition of integral over these generalized configuration
spaces. This is the problem we state and solve in the present
work, by constructing an explicit algorithm for the integration
over a general algebra. Many examples are discussed in order to
illustrate our construction.

\end{quotation}
\newpage
\resection{Introduction}
The very  idea of supersymmetry leads to the possibility of extending
ordinary classical mechanics to more general cases in which ordinary
configuration variables live together with Grassmann variables. More
recently the idea of extending classical mechanics to more general
situations has been further emphasized with the introduction of
quantum groups, non-commutative geometry, etc. In order to quantize
these general theories, one can try two ways: i) the canonical
formalism, ii) the path-integral quantization. In refs.
\cite{berezin,casalbuoni} classical theories involving Grassmann
variables were quantized by using the canonical formalism. But in this
case, also the second possibility can be easily realized by using the
Berezin's rule for integrating over a Grassmann algebra
\cite{berezin2}. It would be desirable to have a  way to perform the
quantization of theories defined in a general algebraic setting. In
this paper we will make a first step toward this construction, that is
we will give general rules allowing the possibility of integrating
over a given algebra. Given these rules, the next step would be the
definition of the path-integral. In order to define the integration
rules we will need some guiding principle. So let us start by
reviewing how the integration over Grassmann variables come about. The
standard argument for the Berezin's rule is translational invariance.
In fact, this guarantees the validity of the quantum action principle.
However, this requirement seems to be too technical and we would
rather prefer to rely on some more  physical argument, as the one
which is automatically satisfied by the path integral representation
of an amplitude, that is the combination law for  probability
amplitudes. This is a simple consequence of the factorization
properties of the functional measure and of the additivity of the
action. In turn, these properties follow in a direct way from the very
construction of the path integral starting from the ordinary quantum
mechanics. We recall that the construction consists in the computation
of the matrix element $\langle q_f,t_f|q_i,t_i\rangle$, ($t_i<t_f$) by
inserting the completeness relation
\be
\int\; dq\;|q,t\rangle\langle q,t|=1
\label{completezza}
\ee
inside the matrix element at the intermediate times $t_a$
($t_i<t_a<t_f$, $a=1,\cdots,N$), and taking the limit $N\to\infty$
(for sake of simplicity we consider here the quantum mechanical case
of a single degree of freedom). The relevant information leading to
the composition law is nothing but the completeness relation
(\ref{completezza}). Therefore we will assume the completeness as the
basic principle to use in order to define the integration rules
 over a generic algebra. In
this paper we will limit our task to the construction of the
integration rules, and we will not do any attempt to construct the
functional integral in the general case. The extension of the relation
(\ref{completezza}) to a configuration space different from the usual
one is far from being trivial. However, we can use an approach that
has been largely used in the study of non-commutative geometry
\cite{connes} and  of quantum groups \cite{drinfeld}. The approach
starts from the observation that in the normal case one can
reconstruct a space from the algebra of its functions . Giving this
fact, one lifts all the necessary properties in the function space and
avoids to work on the space itself. In this way one is able to deal
with cases in which no concrete realization of the space itself
exists. We will see in Section 2 how to extend the relation
(\ref{completezza}) to the algebra of  functions. In Section 3 we will
generalize the considerations of Section 2 to the case of an arbitrary
algebra. In Section 4 we will discuss numerous examples of our
procedure. The approach to the integration on the Grassmann algebra,
starting from the requirement of completeness was discussed long ago
by Martin \cite{martin}.

\resection{The algebra of functions}

Let us consider a quantum dynamical system and an operator having a
complete set of eigenfunctions. For instance one can consider a
one-dimensional free particle. The hamiltonian eigenfunctions are
\be
\psi_k(x)=\frac 1 {\sqrt{2\pi}}\exp{(-ikx)}
\ee
Or we can consider the orbital angular momentum, in which case the
eigenfunctions are the spherical harmonics $Y_\ell^m(\Omega)$. In
general the eigenfunctions satisfy orthogonality relations
\be
\int\psi_n^*(x)\psi_m(x)\;dx=\delta_{nm}
\label{ortogonalita}
\ee
(we will not distinguish here between discrete and continuum
spectrum). However $\psi_n(x)$ is nothing but the representative
in the $\langle x|$ basis of the eigenkets $|n\rangle$ of the
hamiltonian
\be
\psi_n(x)=\langle x| n\rangle
\ee
Therefore the eq. (\ref{ortogonalita}) reads
\be
\int\langle n|x\rangle\langle x|m\rangle\;dx=\delta_{nm}
\ee
which is equivalent to say that the $|x\rangle$ states form a
complete set and that $|n\rangle$  and $|m\rangle$ are orthogonal.
But this means that we can implement the completeness in the
$|x\rangle$ space by means of the orthogonality relation obeyed by
the eigenfunctions defined over this space. Another important
observation is that the orthonormal functions define an algebra.
In fact we can expand the product of two eigenfunctions in terms
of the eigenfunctions themselves
\be
\psi_m(x)\psi_n(x)=\sum_p c_{nmp}\psi_p(x)
\label{algebra}
\ee
with
\be
c_{nmp}=\int\psi_n(x)\psi_m(x)\psi_p^*(x)\;dx
\label{cnmp}
\ee
For instance, in the case of the free particle
\be
c_{kk'k''}=\frac 1 {\sqrt{2\pi}}\delta(k+k'-k'')
\ee
In the case of the angular momentum one has the product formula
\cite{messiah}
\bea
Y_{\ell_1}^{m_1}(\Omega) Y_{\ell_2}^{m_2}(\Omega)&=&
\sum_{L=|\ell_1-\ell_2|}^{\ell_1+\ell_2}\sum_{M=-L}^{+L}\left[
\frac{(2\ell_1+1)(2\ell_2+1)}{4\pi(2L+1)}\right]\nn\\
&\times&\langle\ell_1\ell_2 0 0| L0\rangle\langle \ell_1
\ell_2 m_1 m_2|
LM\rangle Y_L^M(\Omega)
\eea
where $\langle j_1 j_1 m_1 m_2|JM\rangle$ are the Clebsch-Gordan
coefficients. A set of eigenfunctions can then be considered as a
basis of the algebra (\ref{algebra}), with structure constants given
by (\ref{cnmp}). Any function can be expanded in terms of the complete
set $\{\psi_n(x)\}$, and therefore it will be convenient, for the
future, to introduce a generalized Fock space ${\cal F}$ build up in
terms of the eigenfunctions
\be
|\psi\rangle=\left(\matrix{\psi_0(x)\cr\psi_1(x)\cr\cdots\cr\psi_n(x)
\cr\cdots\cr}\right)
\ee
A function $f(x)$ such that
\be
f(x)=\sum_n a_n\psi_n(x)
\ee
 can be represented as
\be
f(x)=\langle a|\psi\rangle
\ee
where
\be
\langle a|=\left(a_0,a_1,\cdots,a_n,\cdots\right)
\ee
To write the orthogonality relation in terms of this new formalism
it is convenient to realize the complex conjugation as a linear
operation on $\FF$. In fact, $\psi_n^*(x)$ itself can be expanded
in terms of $\psi_n(x)$
\be
\psi_n^*(x)=\sum_n\psi_m(x)C_{mn}
\ee
or
\be
|\psi^*\rangle=C^T|\psi\rangle
\label{c_matrix}
\ee
Defining a bra in $\FF$ as the transposed of the ket
$|\psi\rangle$
\be
\langle\psi|=(\psi_0(x),\psi_1(x),\cdots(x),\psi_n(x),\cdots)
\ee
the orthogonality relation becomes
\be
\int|\psi\rangle\langle\psi^*|\;dx=\int |\psi\rangle\langle\psi|C
\;dx=1
\label{newcompleteness}
\ee
Notice that  by taking the complex conjugate of eq.
(\ref{c_matrix}), we get
\be
CC^*=1
\ee
The relation (\ref{newcompleteness}) makes reference only to the
elements of the algebra of functions that we have organized in the
space $\FF$, and it is the key element in order to define the
integration rules on the algebra. In fact, we can now use the algebra
product to reduce the  expression (\ref{newcompleteness}) to a linear
form
\be
\delta_{nm}=\sum_\ell\int \psi_n(x)\psi_\ell(x)C_{\ell m}\; dx=
\sum_{\ell,p}c_{n\ell  p}C_{\ell m}\int \psi_p(x)\; dx
\ee
If the set of equations
\be
\sum_p A_{nmp}\int\psi_p(x)\; dx=\delta_{nm},~~~
A_{nmp}=\sum_\ell c_{n\ell  p}C_{\ell m}
\ee
has a solution for $\int\psi_p(x)\; dx$, then we are able to define
the integration over all the algebra,  by linearity. We will show in
the following that indeed a solution exists for many interesting
cases.  For instance a solution always exists, if the constant
function is in the set $\{\psi_p(x)\}$.
 However we will not try here to define the conditions under
which the equations are satisfied. Let us just show what we get for
the free particle. The matrix $C$ is easily obtained by noticing that
\bea
\left(\frac 1{\sqrt{2\pi}}\exp(-ikx)\right)^*&=&\frac 1{\sqrt{2\pi}}\exp(ikx)\nn\\
&=&\int\;dk'\delta(k+k')\frac 1{\sqrt{2\pi}}\exp(-ik'x)
\eea
and therefore
\be
C_{kk'}=\delta(k+k')
\ee
It follows
\be
A_{kk'k''}=\int\;dq\;\delta(k'+q)\frac 1{\sqrt{2\pi}}\delta(q+k-k'')=
\frac 1{\sqrt{2\pi}}\delta(k-k'-k'')
\ee
from which
\be
\delta(k-k')=\int\;dk''\;\int\;A_{kk'k''}\psi_{k''}(x) dx=
\int\frac 1{2\pi}\exp(-i(k-k')x)dx
\ee
This example is almost trivial, but it shows how, given the structure
constants of the algebra, the property of the exponential of being the
Fourier transform of the delta-function follows automatically from the
formalism. In fact, what we have really done it has been {\bf{ to
define the integration rules}} by using only the algebraic properties
of the exponential. As a result, our integration rules require that
the integral of an exponential is a delta-function. One can perform
similar steps in the case of the spherical harmonics, where the $C$
matrix is given by
\be
C_{(\ell,m),(\ell',m')}=(-1)^m\delta_{\ell,\ell '}\delta_{m,-m'}
\ee
and then using the
constant function $Y_0^0=1/\sqrt{4\pi}$, in the completeness
relation.

 The procedure we have outlined here is the one  that we will
generalize in the next Section to arbitrary algebras. Before doing
that we will consider the possibility of a further generalization. In
the usual path-integral formalism sometimes one makes use of the
coherent states  instead of the position  operator eigenstates. In
this case the basis in which one considers the wave functions is a
basis of eigenfunctions of a non-hermitian operator
\be
\psi(z)=\bra\psi z\rangle
\ee
with
\be
a\ket z=\ket z z
\ee
The wave functions of this type close an algebra, as
$\langle z^*|\psi\rangle$ do. But this time the two types of
eigenfunctions are not connected by any linear operation. In
fact,
the completeness relation is defined on the direct product of the
two algebras
\be
\int\frac{dz^*dz}{2\pi i}\exp(-z^*z)|z\rangle\langle z^*|=1
\ee
Therefore, in similar situations, we will not define the integration
over the original algebra, but rather on the algebra obtained by the
tensor product of the  algebra times a copy. The copy corresponds to
the complex conjugated functions of the previous example.

\resection{Eigenvalues and eigenvectors for a generic
algebra} Let us start with a generic algebra $\AAA$ with $n+1$
elements $x_i$, with $i=0,1,\cdots n$. In the following we will
consider also the case $n\to\infty$. We assume the multiplication
rules
\be
x_i x_j=f_{ijk}x_k
\ee
with the usual convention of sum over the repeated indices.
The structure constants $f_{ijk}$ define uniquely the
algebraic structure. Consider for instance the case of an
abelian algebra. In this case
\be
x_i x_j=x_j x_i\longrightarrow f_{ijk}=f_{jik}
\label{commutativity}
\ee
The associativity condition reads
\be
x_i(x_j x_k)=(x_i x_j) x_k
\ee
leading to the condition
\be
f_{ilm}f_{jkl}=f_{ijl}f_{lkm}
\label{associativity}
\ee
We will make use of this equation in the following. An algebra
being a vector space, the most general function on the algebra
(that is a mapping $\AAA\to\AAA$) is a linear one:
\be
f(a,x)=\sum_{i=0}^n a^i x_i
\ee
Of course, this relation defines a mapping between the $n+1$
dimensional row-vectors and the functions on the algebra, that is
a mapping between $\CC^{n+1}$ and $\AAA$.
\be
\langle a|\equiv (a^0,a^1,\cdots,a^n)\longleftrightarrow f(x_i)
\label{mapping}
\ee
By proceeding as in Section 2 we introduce the space $\FF$ of
vectors build up in terms of the generators of the algebra
\be
| x\rangle=\left(\matrix{x_0\cr x_1\cr\cr \cdot\cr\cdot\cr
x_n\cr}\right),~~~~~~\ket x\in\FF
\ee
The mapping (\ref{mapping}) becomes
\be
\langle a|\longleftrightarrow \langle a| x\rangle
\ee
The action of a linear operator on $\FF$ is
induced by its action on $\CC^{n+1}$
\be
 \langle a'|=\langle a|O \longleftrightarrow  \langle a'| x\rangle
=\langle a|O| x\rangle=\langle a| x'\rangle
\ee
where
\be
 | x'\rangle=O| x\rangle
\ee
In order to be able to generalize properly the discussion of
Section 2 it will be of fundamental importance to
 look for linear operators having
the vectors $|x\rangle$ as eigenvectors and the algebra
elements $x_i$ as eigenvalues. As we shall see this notion
is strictly related to the mathematical concept of {\bf right and
left multiplication algebras} associated to a given algebra.
The linear operators we are looking for are defined by the
relation
\be
X_i |x\rangle=|x\rangle x_i
\label{eigenvalues}
\ee
that is
\be
(X_i)_{jk}x_k=x_jx_i=f_{jik}x_k
\ee
or
\be
(X_i)_{jk}=f_{jik}
\ee
To relate this notion to the right multiplication algebra, let us
consider the right multiplication of an arbitrary element of the
algebra by a generator
\bea
\langle a| x\rangle x_i&=&f(a,x)x_i=\sum_ja^j x_j x_i=
\sum_j a^j f_{jik} x_k\nn\\
&=& \sum_j a^i(X_i)_{jk} x_k=
f(aX_i,x)= \langle a X_i| x\rangle \nn\\
&=&\langle a|(X_i)|
 x\rangle
\eea
from which the (\ref{eigenvalues}) follows. Therefore the matrix $X_i$
corresponds to the linear transformation induced on the algebra by the
right multiplication by the element $x_i$. In a complete analogous way
we can consider column vectors $|b\rangle$, and define a function on
the algebra as
\be
g(x,b)=\langle\tilde
x|b\rangle=\left(\matrix{x_0,x_1,\cdot\cdot,x_n\cr}\right)
\left(\matrix{b^0\cr
b^1\cr\cdot\cr\cdot\cr}\right)=\sum_ix_ib^i
\ee
Now let us consider the left multiplication
\be
x_i g(x,b)=x_i\langle\tilde x|b\rangle =\sum_j f_{ijk}x_k
b^j
\ee
Defining
\be
(\Pi_i)_{kj}=f_{ijk}
\ee
we get
\be
x_i g(x,b)=\sum_j x_k(\Pi_i)_{kj} b^j =g(x,\Pi_i b)=
\langle\tilde x|\Pi_i|b \rangle
\ee
therefore
\be
\langle\tilde x|\Pi_i=x_i\langle\tilde x|
\label{lefteigenvalues}
\ee
The two matrices $X_i$ and $\Pi_i$ corresponding to right
and left multiplication  are generally
different:
\be
(X_i)_{jk}=f_{jik},~~~~~(\Pi_i)_{jk}=f_{ikj}
\ee
In terms of the matrices $X_i$ and $\Pi_i$ one can characterize
different algebras. For instance, consider  the abelian case. It
follows from eq. (\ref{commutativity})
\be
X_i=\Pi_i^T
\ee
If the algebra is associative, then from (\ref{associativity})
the following three relations can be shown to be equivalent:
\be
X_iX_j=f_{ijk}X_k,~~~\Pi_i\Pi_j=f_{ijk}\Pi_k,~~~[X_i,\Pi_j^T]=0
\ee
The first two say that  $X_i$ and  $\Pi_i$ are linear
representations of the algebra. The third that the right and
left multiplication commute for associative algebras.

Recalling the discussion in Section 2 we would like first consider
the case of a basis originating from hermitian operators. Notice
that the generators $x_i$ play here the role of generalized
dynamical variables. It is then natural to look for the case in
which the operators $X_i$ admit both eigenkets and eigenbras. This
will be the case if
\be
\Pi_i=CX_iC^{-1}
\label{matrice_c}
\ee
that is $\Pi_i$ and $X_i$ are connected by a non-singular
$C$ matrix. This matrix is exactly the analogue of the matrix $C$
defined in eq. (\ref{c_matrix}).
 From (\ref{lefteigenvalues}), we get
\be
\langle\tilde x|CX_iC^{-1}=x_i\langle \tilde x|
\ee
By putting
\be
\langle x|=\langle\tilde x|C
\label{conjugation}
\ee
we have
\be
\langle x|X_i=x_i\langle x|
\label{bra}
\ee
In this case, the equations (\ref{eigenvalues}) and
(\ref{bra}) show that $X_i$ is the analogue of an hermitian
operator. We will define now the
integration over the algebra by requiring that
\be
\int_{(x)}|x\rangle\langle x|=1
\ee
where $1$ the identity matrix on the
$(n+1)\times(n+1)$dimensional linear space of the linear
mappings on the algebra. In more explicit terms we get
\be
\int_{(x)}x_i ( x_k C_{kj})=\delta_{ij}
\ee
or
\be
 \int_{(x)}x_i  x_j=(C^{-1})_{ij}
\label{integration1}
\ee
as well as
\be
\int_{(x)}f_{ijk}x_k= (C^{-1})_{ij}
\ee
If we can invert this relation in terms of $\int_{(x)}x_i$, we can say
to have defined the integration over the algebra, because we can
extend the operation by linearity. In particular, if ${\cal A}$ is an
algebra with identity, let us say $x_0=1$, then, by using
(\ref{integration1}), we get
\be
\int_{(x)} x_i=(C^{-1})_{0i}
\ee
and it is always possible to define the integral.

We will discuss now the transformation properties of the integration
measure with respect to an automorphism of the algebra. In particular,
we will restrict our analysis to the case of a simple algebra (that is
an algebra having as ideals only the algebra itself and the null
element). Let us consider an invertible linear transformation on the
basis of the algebra leaving invariant the multiplication rules (that
is an automorphism)
\be
x_i'=S_{ij}x_j
\ee
with
\be
x_i'x_j'=f_{ijk}x_k'
\ee
This implies the following conditions for the transformation
$S$
\be
S_{il}S_{jm}f_{lmp}=f_{ijk}S_{kp}
\ee
This relation can be written in a more convenient form in terms of the
matrices $X_i$ and $\Pi_i$
\be
SX_iS^{-1}=(S^{-1})_{ij}X_j,~~~S^{T-1}\Pi_i
S^T=(S^{-1})_{ij}\Pi_j
\ee
In the case we are considering here $X_i$ and $\Pi_i$ are related by
the $C$ matrix (see eq. (\ref{matrice_c}), and therefore we get
\be
(C^{-1}S^{T-1}C)X_i(C^{-1}S^TC)=SX_iS^{-1}
\ee
For a simple algebra, one can show that the enveloping algebra of the
right and left multiplications forms an irreducible set of linear
operators \cite{schafer}, and therefore by the Shur's lemma we obtain
\be
C^{-1}S^T C=kS^{-1}
\ee
where $k$ is a constant.
It follows
\be
\bra x\to \bra {\tilde x} S^TC=k\bra{\tilde x}CS^{-1}=k\bra x
S^{-1}
\ee
Now we require
\be
\int_{(x')}\ket {x'}\bra {x'}=  \int_{(x)}\ket x\bra x
\ee
which is satisfied by taking
\be
 \int_{(x')} =\frac 1 k  \int_{(x)}
 \label{change}
\ee
In fact
\be
\int_{(x')} \ket {x'}\bra {x'}=\frac 1 k \int_{(x)} S\ket {x}k\bra
{x}S^{-1}=1
\ee
Let us consider now the case in which the automorphism $S$ can be
exponentiated in the form
\be
S=\exp(\alpha D)
\ee
Then $D$ is a derivation of the algebra, as it follows from
\be
(x_i x_j)'=x_i'x_j'\rightarrow \exp(\alpha D)(x_ix_j)=
(\exp(\alpha D) x_i)(\exp(\alpha D)x_j)
\ee
by taking $\alpha$ infinitesimal. If it happens that for this
particular automorphism $S$, one has $k=1$, the integration measure is
invariant (see eq. (\ref{change})). Then, the integral satisfies
\be
\int_{(x)}D(f(x))=0
\label{byparts}
\ee
for any function $f(x)$ on the algebra. On the contrary,  a derivation
always defines an automorphism of the algebra by exponentiation. So,
if the corresponding $k$ is equal to one, the equation (\ref{byparts})
is always valid.

Of course it may happen that the  $C$ matrix does not exist. This
would correspond to the case of  non-hermitian operators discussed in
Section 2. So we look for a copy $\AAA^*$ of the algebra. By calling
$x^*$ the elements of $\AAA^*$, the corresponding generators will
satisfy
\be
x_i^*x_j^*=f_{ijk}x_k^*
\ee
It follows
\be
\langle{\tilde x}^*|\Pi_i=x_i^*\langle{\tilde x}^*|
\ee
Then, we define the integration rules on the tensor product
of $\AAA$ and $\AAA^*$ in such a way that the
completeness relation holds
\be
\int_{(x,x^*)}|x\rangle\langle{\tilde x}^*|=1
\ee
or
\be
\int_{(x,x^*)}x_ix_j^*=\delta_{ij}
\label{integration2}
\ee
This second type of integration is invariant under orthogonal
transformation or unitary transformations, according to the way in
which the $^*$ operation acts on the transformation matrix $S$. If
$^*$ acts on complex numbers as the ordinary conjugation, then we have
invariance under unitary transformations, otherwise if $^*$ leaves
complex numbers invariant, then the invariance is under orthogonal
transformations. Notice that the invariance property does not depend
on $S$ being an automorphism of the original algebra or not.

The two cases considered here are not mutually exclusive.
In fact, there are situations that can be analyzed from both
points of view.

We  want also to emphasize that this approach  does not pretend to be
complete and that we are not going to give any theorem about the
classification of the algebras with respect to the integration. What
we are giving is rather a set of rules that one can try to apply in
order to define an integration over an algebra. As argued before,
there are algebras that do not admit the integration as we have
defined in (\ref{integration1}) or in (\ref{integration2}). Consider,
for instance, a simple Lie algebra. In this case we have the relation
\be
f_{ijk}=f_{jki}
\ee
which implies
\be
X_i=\Pi_i
\ee
or $C=1$. Then the eq. (\ref{integration1}) requires
\be
\delta_{ij}=\int_{(x)}x_ix_j=\int_{(x)}f_{ijk}x_k
\ee
which cannot be satisfied due to the antisymmetry of the
structure constants. Therefore, we can say that, according to
our integration rules, there are algebras with a complete set of
states and algebras which are not complete.

\resection{Examples}

In this Section we will discuss several examples of both
types of integration.
\subsection{The bosonic case}

We will start trying to reproduce the integration rules in the
bosonic case. It is convenient to work in the coherent state
basis. The coherent states are defined by the relation
\be
a|z\rangle=|z\rangle z
\ee
where $a$ is the annihilation operator, $[a,a^\dagger]=1$. The
representative of a state at fixed occupation number in
the coherent state basis is
\be
\langle n| z\rangle=\frac{z^n}{\sqrt{n!}}
\ee
So we will consider as elements of the algebra the quantities
\be
x_i=  \frac{z^i}{\sqrt{i!}},~~~ i=0,1,\cdots,\infty
\ee
The states in $\FF$ are therefore
\be
\left(\matrix{1\cr  z \cr\dd{ {z^2}/{\sqrt{2!}}}
\cr\cdot\cr\cdot\cr}\right)
\ee
The algebra is defined by the multiplication rules
\be
x_ix_j=\frac{z^{i+j}}{\sqrt{i!\,j!}}=x_{i+j}
\sqrt{\frac{(i+j)!}{i!\,j!}}
\ee
from which
\be
f_{ijk}=\delta_{i+j,k}\sqrt{\frac{k!}{i!\,j!}}
\ee
It follows
\be
(X_i)_{jk}=\delta_{i+j,k}\sqrt{\frac{k!}{i!\,j!}}
\ee
and
\be
(\Pi_i)_{jk}=\delta_{i+k,j}\sqrt{\frac{j!}{i!\,k!}}
\ee
In particular we get
\be
(X_1)_{jk}=\sqrt{k}\,\delta_{j+1,k},~~~~
(\Pi_1)_{jk}=\sqrt{k+1}\,\delta_{j-1,k}
\ee
Therefore $X_1$ and $\Pi_1$ are nothing but the representative,
in the occupation number basis, of the annihilation and creation
operators respectively. It follows that the $C$ matrix cannot
exist, because $[X_1,\Pi_1]=1$, and a unitary transformation
cannot change this commutation relation into $[\Pi_1,X_1]=-1$.
For an explicit proof consider, for instance, $X_1$
\be
X_1=\left(\matrix{0 & \sqrt{1} & 0 & 0 & \cdots\cr
                  0 & 0 & \sqrt{2} & 0 & \cdots\cr
                  0 & 0 & 0 & \sqrt{3} & \cdots\cr
 \cdot & \cdot &\cdot&\cdot&\cdots\cr
 \cdot & \cdot &\cdot&\cdot&\cdots\cr}\right)
\ee
 If the matrix $C$ would exist
it would be possible to find states $\langle z|$ such that
\be
\langle z|X_1=z\langle z|
\ee
with
\be
\langle z|=(f_0(z),f_1(z),\cdots)
\ee
This would mean
\be
\langle z|X_1=(0,f_0(z),\sqrt{2}f_1(z),\cdots)=
(zf_0(z),zf_1(z),zf_2(z),
\cdots)
\ee
which implies
\be
f_0(z)=f_1(z)=f_2(z)=\cdots=0
\ee
Now, having shown that no $C$
matrix exists, we will consider the complex conjugated algebra
with generators constructed in terms of $z^*$, where
$z^*$ is the complex conjugate of $z$. Then the equation
\be
\langle {\tilde z}^*|\Pi_i= z_i^*\langle {\tilde z}^*|
\ee
is satisfied by
\be
\langle {\tilde z}^*|=(1,\frac {z^*}{\sqrt{1!}},\frac{{z^*}^2}
{\sqrt{2!}},\cdots)
\ee
and the integration rules give
\be
\int_{(z,z^*)}\frac{z^i{z^*}^j}{\sqrt{i!j!}}=\delta_{i,j}
\ee
We see that our integration rules are equivalent to the gaussian
integration
\be
\int_{(z,z^*)}=\int\frac{dz^*dz}{2\pi i}\exp(-|z|^2)
\label{gaussian}
\ee

Another interesting example is again the algebra of multiplication
of the complex numbers but now defined also for negative integer
powers
\be
z^nz^m=z^{n+m},~~~-\infty\le n,m\le +\infty
\ee
with $z$ restricted to the unit circle
\be
z^*=z^{-1}
\ee
Defining the vectors in $\FF$ as
\be
\ket z=\left(\matrix{\cdot\cr z^{-i}\cr\cdot\cr 1\cr
z\cr\cdot\cr z^i\cr\cdot\cr}\right)
\ee
the $X_i$ and $\Pi_i$ matrices are given by
\be
(X_i)_{ij}=\delta_{i+j,k},~~~(\Pi_i)_{ij}=\delta_{i+k,j}
\ee
and now we can construct a $C$ matrix connecting these two set
of matrices. This is easier seen by looking for a bra which is
eigenvector of $X_i$
\be
\bra z X_i=z^i\bra z
\ee
In components, by putting
\be
\bra z=(\cdots, f_{-i}(z),\cdots,f_0(z),\cdots,f_i(z),\cdots)
\ee
 we get
\be
f_j(z)(X_i)_{jk}=f_j(z)\delta_{i+j,k}=f_{k-i}(z)=z^i f_k(z)
\ee
This equation has the solution
\be
f_i(z)=z^{-i}
\ee
therefore
\be
\bra z= (\cdots, z^{i},\cdots,1,\cdots,z^{-i},\cdots)
\ee
The matrix $C$ is given by
\be
(C)_{ij}=(C^{-1})_{ij}=\delta_{i,-j}
\ee
or. more explicitly by
\be
C=\left(\matrix{\cdot & \cdot & \cdot & \cdot & \cdot \cr
                \cdot &   0   &    0  &   1   & \cdot \cr
                \cdot &   0   &    1  &   0   & \cdot \cr
                \cdot &   1   &    0  &   0   & \cdot \cr
                \cdot & \cdot & \cdot & \cdot & \cdot \cr }\right)
\ee
In fact
\be
(C\Pi_i C^{-1})_{lp}=\delta_{l,-m}\delta_{i+n,m}\delta_{n,-p}=
\delta_{i-p,-l}=\delta_{i+l,p}=(X_i)_{lp}
\ee
Notice that  the $C$ matrix is nothing but the
 representation in $\FF$ of
the complex conjugation ($z\to z^*=z^{-1}$).
 The completeness relation reads now
\be
\int_{(z)}z^iz^{-j}=\delta_{ij}
\ee
from which
\be
\int_{(z)}z^k=\delta_{k0}
\ee
 Our algebraic definition of integral can be interpreted as
an integral along a circle $C$ around the origin. In fact we have
\be
\int_{(z)}=\frac{1}{2\pi i}\int_C\frac{dz}{z}
\ee

\subsection{The $q$-oscillator}

A generalization of the bosonic oscillator is the
$q$-bosonic oscillator \cite{biedenharn}. We will use
the definition given in \cite{baulieu}
\be
b\bar b-q\bar b b=1
\ee
with $q$ real and positive.
We assume as elements of the algebra
${\cal A}$, the quantities
\be
x_i=\frac {z^i}{\sqrt{i_q!}}
\ee
where $z$ is a complex number,
\be
i_q=\frac {q^i-1}{q-1}
\ee
and
\be
i_q!=i_q(i-1)_q\cdots 1
\ee
The structure constants are
\be
f_{ijk}=\delta_{i+j,k}\sqrt{\frac{k_q!}{i_q!j_q!}}
\ee
and therefore
\be
(X_i)_{jk}=\delta_{i+j,k}\sqrt{\frac{k_q!}{i_q!j_q!}},~~~~
(\Pi_i)_{jk}=\delta_{i+k,j}\sqrt{\frac{j_q!}{i_q!k_q!}}
\ee
In particular
\be
(X_1)_{jk}=\delta_{j+1,k}\sqrt{k_q}, ~~~~~
(\Pi_1)_{jk}=\delta_{j-1,k}\sqrt{(k+1)_q}
\ee
We see that $X_1$ and $\Pi_1$ satisfy the $q$-bosonic
algebra
\be
X_1\Pi_1-q\Pi_1 X_1=1
\ee
For $q$ real and positive, no $C$ matrix exists, so,
according to our rules
\be
\int_{(z,z^*)_q}\frac{z^i{z^*}^j}{i_q!j_q!}=\delta_{ij}
\ee
This integration can be expressed in terms of the so
called $q$-integral (see ref. \cite{koorwinder}), by
using the representation of $n_q!$ as a $q$-integral
\be
n_q!=\int_0^{1/(1-q)}d_qt\; e_{1/q}^{-qt}\; t^n
\ee
where the $q$-exponential is defined by
\be
e_{q}^{t}=\sum_{n=0}^\infty\frac{z^n}{n_q!}
\ee
and the $q$-integral through
\be
\int_0^a d_qt f(t)=a(1-q)\sum_{n=0}^\infty f(aq^n)q^n
\ee
 Then the two integrations are related by ($z=|z|\exp(i\phi)$)
 \be
 \int_{(z,z^*)_q}=\int\frac{d\phi}{2\pi}\int d_q(|z|^2)\;
 e_{1/q}^{-q|z|^2}
 \ee
 \newpage

\subsection{The fermionic case}

We will discuss now the case of the Grassmann algebra ${\cal G}_1$,
with generators $1,\theta$, such that $\theta^2=0$. The multiplication
rules are
\be
\theta^i\theta^j=\theta^{i+j},~~~ i,j,i+j=0, 1
\label{grassmannrules}
\ee
and zero otherwise (see Table 1).
\vspace{0.4cm}
\begin{center}
\begin{tabular}{|c||c|c|}
\hline
 & 1 & $\theta$\\
\hline\hline
1 & 1 & $\theta$\\
\hline
$\theta $&$\theta $& 0\\
\hline
\end{tabular}
\end{center}
\begin{center}
 Table 1: {\it Multiplication table for the Grassmann algebra ${\cal G}_1$}.
\end{center}
\vspace{0.4cm}
From the multiplication rules we get the structure constants
\be
f_{ijk}=\delta_{i+j,k},~~~~i,~j,~k=0,1
\ee
from which  the explicit expressions for the matrices
$X_i$ and $\Pi_i$ follow
\bea
(X_0)_{ij}&=&f_{i0j}=\delta_{i,j}=\left(\matrix{ 1 & 0\cr
                                               0 & 1\cr}\right)\nn\\
(X_1)_{ij}&=&f_{i1j}=\delta_{i+1,j}=\left(\matrix{ 0 & 1\cr
                                               0 & 0\cr}\right)\nn\\
(\Pi_0)_{ij}&=&f_{0ji}=\delta_{i,j}=\left(\matrix{ 1 & 0\cr
                                               0 & 1\cr}\right)\nn\\
(\Pi_1)_{ij}&=&f_{1ji}=\delta_{i,j+1}=\left(\matrix{ 0 & 0\cr
                                               1 & 0\cr}\right)
\eea
Notice that $X_1$ and $\Pi_1$ are nothing but the ordinary
annihilation and creation Fermi operators with respect to the
vacuum state $\ket 0=(1,0)$.
 The $C$ matrix exists and it is given by
\be
(C)_{ij}=\delta_{i+j,1}= \left(\matrix{ 0 & 1\cr
                  1 & 0\cr}\right)
\ee
The ket and the bra eigenvectors of $X_i$ are
\be
\ket\theta=\left(\matrix{1\cr\theta\cr}\right),~~~~
\bra\theta=(\theta,1)
\label{g1ketandbra}
\ee
and the completeness reads
\be
\int_{{\cal G}_1}\ket\theta\bra\theta=
\int_{{\cal G}_1}\left(\matrix{\theta & 1\cr
                                  0   & \theta\cr}\right)=
\left(\matrix{1 & 0\cr 0 & 1\cr}\right)
\ee
or
\be
\int_{{\cal G}_1}\theta^i\theta^{1-j}=\delta_{i,j}
\ee
which means
\be
\int_{{\cal G}_1}\, 1=0,~~~~\int_{{\cal G}_1}\,\theta=1
\label{grassmannintegr}
\ee

The case of a Grassmann algebra ${\cal G}_n$, which consists of $2^n$
elements obtained by $n$ anticommuting generators
$\theta_1,\theta_2,\cdots,\theta_n$, the identity, $1$, and by all
their products, can be treated in a very similar way. In fact, this
algebra can be obtained by taking a convenient tensor product of $n$
Grassmann algebras ${\cal G}_1$, which means that the  eigenvectors of
the algebra of the left and right multiplications are obtained by
tensor product of the eigenvectors of eq. (\ref{g1ketandbra}). The
integration rules extended by the tensor product give
\be
\int_{{\cal G}_n}\theta_1\theta_2\cdots\theta_n=1
\ee
and zero for all the other cases,
which is equivalent to require for each copy of ${\cal G}_1$
the equations (\ref{grassmannintegr}).
It is worth to mention the case of the Grassmann algebra
${\cal G}_2$ because it can be obtained
by tensor product of ${\cal G}_1$ times a copy
${\cal G}_1^*$. Then we can apply our second method of getting the
integration rules and show that they lead to the same result
with a convenient interpretation of the measure. The algebra
${\cal G}_2$ is generated by $\theta_1,\theta_2$. An involution
of the algebra is given by the mapping
\be
^*:~~~~\theta_1\leftrightarrow \theta_2
\ee
with the further rule that by taking the $^*$ of a product one has to
exchange the order of the factors. It will be convenient to put
$\theta_1=\theta$, $\theta_2=\theta^*$. This allows us to consider
${\cal G}_2$ as ${\cal G}_1\otimes {\cal G}_1^*\equiv ({{\cal
G}_1,^*})$. Then the ket and bra eigenvectors of left and right
multiplication in ${\cal G}_1$ and ${\cal G}_1^*$ respectively are
given by
\be
\ket\theta=\left(\matrix{1\cr\theta\cr}\right),~~~~
\bra{{\tilde\theta}^*}=(1,\theta^*)
\ee
with
\be
\bra{{\tilde\theta}^*}\Pi_i={\theta^*}^i\bra{{\tilde\theta}^*}
\ee
The completeness relation  reads

\be
\int_{({{\cal G}_1},^*)}\ket\theta\bra{{\tilde\theta}^*}=
\int_{({{\cal G}_1},^*)}\left(\matrix{1 & \theta^*\cr
                                  \theta   & \theta\theta^*\cr}\right)=
\left(\matrix{1 & 0\cr 0 & 1\cr}\right)
\ee
This implies
\bea
\int_{({{\cal G}_1},^*)} 1&=&\int_{({{\cal
G}_1},^*)}\theta\theta^*=1\nn\\
\int_{({{\cal G}_1},^*)}\theta&=&\int_{({{\cal
G}_1},^*)}\theta^*=0
\eea
These relations are equivalent to the integration over
${\cal G}_2$ if we do the following identification
\be
\int_{({{\cal G}_1},^*)}=\int_{{{\cal
G}_2}}\exp(-\theta^*\theta)
\ee
Notice that the factor $\exp(-\theta^*\theta)$ plays the same role of
the factor $\exp(-|z|^2)$ appearing in the gaussian measure (eq.
(\ref{gaussian})). In fact it has the same origin,  it comes out of
the norm
\be
\bra{{\tilde\theta}^*} \theta\rangle=
1+\theta^*\theta=\exp(\theta^*\theta)
\ee

\subsection{The case of parastatistics}

 We will discuss now the case of a paragrassmann algebra of order $p$,
${\cal G}^p_1$, with generators $1$, and $\theta$, such that
$\theta^{p+1}=0$. The multiplication rules are defined by
\be
\theta^i\theta^j=\theta^{i+j},~~~ i,j,i+j=0,\cdots, p
\label{paragrassmannrules}
\ee
and zero otherwise (see Table 2).

From the multiplication rules we get the structure constants
\be
f_{ijk}=\delta_{i+j,k},~~~~i,~j,~k=0,1,\cdots,p
\ee
from which we obtain the following expressions for the
matrices $X_i$ and $\Pi_i$:
\be
(X_i)_{jk}=\delta_{i+j,k},~~~~(\Pi_i)_{jk}=\delta_{i+k,j},~~~
i,j,k=0,1\cdots,p
\ee
\vspace{0.4cm}
\begin{center}
\begin{tabular}{|c||c|c|c|c|c|}
\hline
 & 1 & $~~\theta~~$ &~~~$\cdot$~~~& $\theta^{p-1}$ &~~$\theta^p$~~\\
\hline\hline
1 & 1 & $\theta$ & $\cdot$ &$\theta^{p-1}$ & $\theta^p$\\
\hline
$\theta $&$\theta$ &$\theta^2 $&$\cdot$  &$\theta^p$ & 0\\
\hline
$\cdot$ & $\cdot$ &  $\cdot$ & $\cdot$ & $\cdot$ & $\cdot $\\
\hline
$\theta^{p-1}$& $\theta^{p-1}$ &$\theta^p$ & 0 & 0 & 0\\
\hline
$\theta^p$ & $\theta^p$ & 0 & 0 & 0 & 0\\
\hline
\end{tabular}
\end{center}
\begin{center}
Table 2: {\it Multiplication table for the paragrassmann algebra
${\cal G}_1^p$.}
\end{center}
\vspace{0.4cm}

In analogy with the Grassmann algebra we can construct the $C$ matrix
\be
(C)_{ij}=\delta_{i+j,p}
\ee
In fact
\be
(C X_i C^{-1})_{lq}=\delta_{l+m,p}\delta_{i+m,n}\delta_{n+q,p}=
\delta_{i+p-l,p-q}=\delta_{i+q,l}=(\Pi_i)_{lq}
\ee
The ket and the bra eigenvectors of $X_i$ are  given by
\be
\ket\theta=\left(\matrix{1\cr\theta\cr\cdot\cr\theta^p\cr}
\right),~~~~
\bra\theta=(\theta^p,\cdots,\theta,1)
\ee
and the completeness reads
\be
\int_{{\cal G}^p_1}\theta^i\theta^{p-j}=\delta_{ij}
\ee
which means
\be
\int_{{\cal G}^p_1}\, 1=\int_{{\cal G}^p_1}\,\theta=
\int_{{\cal G}^p_1}\,\theta^{p-1}=0
\ee
\be
\int_{{\cal G}^p_1}\,\theta^p=1
\ee
in agreement with the results of ref. \cite{martin} (see also
\cite{isaev}).

\subsection{The algebra of  quaternions}

The quaternionic algebra is defined by the multiplication
rules
\be
e_A e_B=-\delta_{AB}+\epsilon_{ABC}e_C, ~~~ A,B,C=1,2,3
\ee
where $\epsilon_{ABC}$ is the Ricci symbol in 3
dimensions. The quaternions can be realized in terms of the
Pauli matrices $e_A=-i\sigma_A$. The automorphism group of
the quaternionic algebra is $SO(3)$, but it is more useful to
work in the so called split basis
\bea
u_0=\frac 1 2 (1+ie_3),&&u_0^*=\frac 1 2 (1-ie_3)\nn\\
u_+=\frac 1 2 (e_1+ie_{2}),&&u_-=\frac 1 2 (e_1-ie_{2})
\eea
In this basis the multiplication rules are given in Table 3.
\vspace{0.4cm}
\begin{center}
\begin{tabular}{|c||c|c|c|c|}
\hline
 &~~~$u_0$~~~  &~~~ $u_0^*$~~~ &$u_+$& $u_-$ \\
\hline\hline
$u_0$ &$u_0$ & $0$ & $u_+$ &$0$ \\
\hline
$u_0^*$ &$0$ &$u_0^* $&$0$  &$u_-$ \\
\hline
$u_+$ & $0$ &  $u_+$ &~ $ 0 $~ & $-u_0$\\
\hline
$u_-$& $u_-$ &$0$ & $-u_0^*$ &~ $ 0 $~\\
\hline
\end{tabular}
\end{center}
\begin{center}
Table 3: {\it Multiplication table for the  quaternionic algebra.}
\end{center}
\vspace{0.4cm}
The automorphism group of the split basis is $U(1)$, with $u_0$ and
$u_0^*$ invariant and $u_+$ and $u_-$ with charges $+1$ and $-1$
respectively. The vectors in $\FF$ are
\be
\ket u=\left(\matrix{u_0\cr u_0^*\cr u_+\cr u_-\cr}\right)
\ee
The matrices $X_A$ and $\Pi_A$ satisfy the quaternionic algebra
because this  is an associative algebra. So $X_+$ and  $X_-$ satisfy
the algebra of a Fermi oscillator (apart a sign). It is easy to get
explicit expressions for the left and right multiplication matrices
and check that the $C$ matrix exists and that it is given by
\be
C=\left(\matrix{1 & 0 & 0 & 0\cr
        0 & 1 & 0 & 0\cr
        0 & 0 & 0 & -1\cr
        0 & 0 & -1 &0\cr}\right)
\ee
Therefore
\be
\bra u=\left(u_0,u_0^*,-u_-,-u_+\right)
\ee
The exterior product is given by
\bea
\ket u\bra u&=&\left(\matrix{u_0\cr u_0^*\cr u_+\cr u_-}\right)
\left(u_0,u_0^*,-u_-,-u_+\right)\nn\\&=&
\left(\matrix{ u_0 & 0 & 0 & -u_+\cr
                0  & u_0^* & -u_- & 0\cr
                0  & u_+ &  u_0 & 0\cr
                u_- & 0 & 0 &
 u_0^*\cr}
\right)
\eea
According to our integration rules we get
\be
\int_{(u)}\,u_0=\int_{(u)}\,u_0^*=1,~~~~
\int_{(u)}\,u_+=\int_{(u)}\,u_-=0
\ee
In terms of the original basis for the quaternions we get
\be
\int_{(u)}1=2,~~~~\int_{(u)}e_A=0
\ee
and we see that, not unexpectedly, the integration coincides with
taking the trace in the $2\times 2$ representation of the
quaternions. That is, given an arbitrary functions $f(u)$ on the
quaternions we get
\be
\int_{(u)} f(u)=Tr[f(u)]
\ee
By considering the scalar product
\be
\langle u'|u\rangle=u_0'u_0+{u_0^*}'u_0-{u_-^*}'u_+-u_+'u_-^*
\ee
we see that
\be
\langle u|u\rangle=2
\ee
and
\be
\int_{(u)}\langle u'|u\rangle= u_0'+{u_0^*}'=1
\ee
Therefore $\langle u'|u\rangle$ behaves like a delta-function.

\subsection{The algebra of octonions}

We will discuss  now how to integrate over the octonionic algebra (see
\cite{ottonioni}). This algebra (said also a Cayley algebra) is
defined in terms of the multiplication table of its seven imaginary
units $e_A$
\be
e_Ae_B=-\delta_{AB}+ a_{ABC}e_C,~~~A,B,C=1,\cdots,7
\ee
where $a_{ABC}$ is completely antisymmetric and equal to +1 for
$(ABC)=(1,2,3)$, (2,4,6), (4,3,5), (3,6,7), (6,5,1), (5,7,2) and
(7,1,4). The automorphism group of the algebra is $G_2$. We define
also in this case the split basis as
\bea
u_0=\frac 1 2 (1+ie_7),&&u_0^*=\frac 1 2 (1-ie_7)\nn\\
u_i=\frac 1 2 (e_i+ie_{i+3}),&&u_i^*=\frac 1 2 (e_i-ie_{i+3})
\eea
where $i=1,2,3$. In this basis  the  multiplication rules are given in
Table 4.

\vspace{0.4cm}
\begin{center}
\begin{tabular}{|c||c|c|c|c|}
\hline
 &~~~$u_0$~~~  &~~~ $u_0^*$~~~ &$u_j$& $u_j^*$ \\
\hline\hline
$u_0$ &$u_0$ & $0$ & $u_j$ &$0$ \\
\hline
$u_0^*$ &$0$ &$u_0^* $&$0$  &$u_j^*$ \\
\hline
$u_i$ & $0$ &  $u_i$ &~ $\epsilon_{ijk}u_k^*$~ & $-\delta_{ij}u_0$\\
\hline
$u_i^*$& $u_i^*$ &$0$ & $-\delta_{ij}u_0^*$ &~ $\epsilon_{ijk} u_k$~\\
\hline
\end{tabular}
\end{center}
\begin{center}
Table 4: {\it Multiplication table for the  octonionic algebra.}
\end{center}
\vspace{0.4cm}
\vspace{0.4cm}
This algebra is non-associative and in the split basis it has an
automorphism group $SU(3)$. The non-associativity can be checked by
taking, for instance,
\be
u_i(u_j u^*_k)=u_i(-\delta_{jk} u_0)=0
\ee
and comparing with
\be
(u_i u_j)u^*_k=\epsilon_{ijm}u_m^*
u^*_k=-\epsilon_{ijk}\epsilon_{kmn}u_n
\ee
The vectors in $\FF$ are
\be
\ket u=\left(\matrix{u_0\cr u_0^*\cr u_i\cr u_i^*\cr}\right)
\ee
and one can easily evaluate the matrices $X$ and $\Pi$
corresponding to right and left multiplication.
We will not
give here the explicit expressions, but one can easily see some
properties. For instance, one can evaluate the anticommutator
$[X_i,X^*_j]_+$, by using the following relation
\be
[X_i,X^*_j]_+\ket u=X_i\ket u u_j^*+X_j^*\ket u u_i=
(\ket u u_i)u_j^*+(\ket u u_j^*) u_i
\ee
The algebra of the anticommutators of $X_i,X_i^*$ turns out to
be the algebra of three Fermi oscillators (apart from the
sign)
\be
[X_i,X^*_j]_+=-\delta_{ij},~~~ [X_i,X_j]_+=0,~~~[X_i^*,X^*_j]_+=0
\ee
The matrices $X_0$ and $X_0^*$ define orthogonal projectors
\be
X_0^2=X_0,~~~ (X_0^*)^2=X_0^*,X_0X_0^*=X_0^*X_0=0
\ee
Further properties are
\be
X_0+X_0^*=1
\ee
and
\be
X_i^*=-X_i^T
\ee
Similar properties hold for the left multiplication matrices.
One can also show that there is a matrix $C$ connecting left and
right multiplication matrices. This is given by
\be
C=\left(\matrix{1 & 0 & 0 & 0\cr
        0 & 1 & 0 & 0\cr
        0 & 0 & 0 & -1_3\cr
        0 & 0 & -1_3 &0\cr}\right)
\ee
where $1_3$ is the $3\times 3$ identity matrix. It follows that the
eigenbras of the matrices of type $X$, are
\be
\bra u=\left(u_0,u_0^*,-u_i^*,-u_i\right)
\ee
For getting the integration rules we need the external product
\bea
\ket u\bra u&=&\left(\matrix{u_0\cr u_0^*\cr u_i\cr u_i^*}\right)
\left(u_0,u_0^*,-u_j^*,-u_j\right)\nn\\&=&
\left(\matrix{ u_0 & 0 & 0 & -u_j\cr
                0  & u_0^* & -u_j^* & 0\cr
                0  & u_i & \delta_{ij} u_0 & -\epsilon_{ijk} u_k^*\cr
                u_i^* & 0 & -\epsilon_{ijk} u_k &
                     \delta_{ij} u_0^*\cr}
\right)
\eea
According to our rules we get
\be
\int_{(u)}\,u_0=\int_{(u)}\,u_0^*=1,~~~~
\int_{(u)}\,u_i=\int_{(u)}\,u_i^*=0
\ee
Other interesting properties are
\be
\langle u\ket u=u_0+u_0^*+3u_0^*+3u_0=4
\ee
and using
\be
\langle u'\ket u=u_0'u_0+{u_0^*}'u_0-{u_i^*}'u_i-u_i'u_i^*
\ee
we get
\be
\int_{(u)}\langle u'\ket u=u_0'+{u_0^*}'=1
\ee
Showing that $\langle u'\ket u$ behaves like a delta-function.
\newpage
\resection{Conclusions and outlook}

In this paper we have shown how it is possible to define an integral
over an arbitrary algebra. The main idea is to restate the
completeness relation in the configuration space (the space spanned by
the eigenkets $\ket x$ of the position operator), in terms of the wave
functions (the functions on the configuration space). In this way the
completeness relation can be understood in algebraic terms and this
has allowed us to define the integration rules over an arbitrary
algebra in terms of the completeness relation itself. The physical
motivation  to require the completeness relation is that it ensures
the composition law for probabilities, as discussed in the
Introduction.

The motivations of the present work come from searching a way of
quantizing a theory defined on a configuration space made up of
non-commuting variables, the simplest example being the case of
supersymmetry. The work presented here is only a first approach to
this subject. First of all we have limited our investigation to the
construction of the integration rules, but we have not tried to study
under which conditions they are satisfied in a given algebra. Or, said
in a different way, we have not looked for a classification of
algebras with respect to the integration rules we have defined.
Second, in order to build up the functional integral, a further step
is necessary. One needs a different copy of the given algebra to each
different time along the path-integration. This should be done by
taking convenient tensor products of copies of the algebra. Given
these limitations, we think, however, that the step realized in this
work is  a necessary one in order to solve the problem of quantizing
the general theories discussed here.


\end{document}